\documentclass[11pt,a4paper]{article}
\usepackage{jheppub}
\usepackage[T1]{fontenc}
\usepackage{xcolor}
\usepackage{graphicx}
\usepackage{bm}
\usepackage{amsmath,amsfonts,amssymb}
\usepackage{mathtools}
\usepackage{braket}
\usepackage[normalem]{ulem}
\usepackage{setspace} 
\usepackage{here}
 
\begin{document}


\title{Hill-top inflation from Dai-Freed anomaly in the standard model-- A solution to the iso-curvature problem of the axion dark matter--}
\author[a,b]{Masahiro Kawasaki,}
\author[c,b]{Tsutomu T. Yanagida}
\affiliation[a]{ICRR, University of Tokyo, Kashiwa, 277-8582, Japan}
\affiliation[b]{Kavli IPMU (WPI), UTIAS, University of Tokyo, Kashiwa, 277-8583, Japan}
\affiliation[c]{Tsung-Dao Lee Institute and School of Physics and Astronomy, Shanghai Jiao Tong University, 520 Shengrong Road, Shanghai, 201210, China}
\abstract{
The discrete symmetry $Z_4$ in the standard model (SM)  with three right-handed neutrinos is free from the Dai-Freed anomaly. 
Motivated by this $Z_4$ symmetry, we constructed a topological inflation model consistent with all known constraints and observations. 
However, we assumed a specific inflaton potential in the previous work. 
In this paper we extend the inflaton potential in a more general form allowed by the discrete $Z_4$ gauge symmetry and show that consistent hilltop inflation is realized.
We find that the Hubble parameter $H_\mathrm{inf}$ can be smaller than $\simeq 10^{9}$ GeV so that the isocurvature fluctuations of the axion dark matter are sufficiently suppressed.
Furthermore, the running of the spectral index can be as large as $dn_s/\ln k  \simeq  0.0018$ which will be tested in future CMB observations. 
Since this discrete $Z_4$ acts on the SM, the inflaton can couple to pairs of the right-handed neutrinos and hence the reheating temperature can be high as $\sim 10^{10}$ GeV, producing the cosmic baryon asymmetry naturally through the thermal leptogenesis.

}

\emailAdd{kawasaki@icrr.u-tokyo.ac.jp}
\emailAdd{tsutomu.tyanagida@sjtu.edu.cn}

\maketitle

\section{Introduction}

Anomalies give strong constraints on models based on quantum field theories. If there are anomalies of gauge symmetries, they must be canceled out to have consistent gauge symmetries in quantum field theories. It is known that the standard model (SM) is free not only from the standard gauge anomalies \cite{Adler:1969gk,Bell:1969ts}, but also from the anomalies of large gauge transformations called  Dai-Freed anomalies \cite{Dai:1994kq,Witten:2015aba} (see also \cite{Yonekura:2016wuc}).

However, if we impose additional discrete symmetries in the SM, it is not necessarily anomaly free and we should add new fermions to cancel the anomalies in the SM. 
In fact, if we impose a discrete $Z_4$ gauge symmetry in the SM, the theory has the Dai-Freed anomalies. 
The introduction of one right-handed neutrino for each generation, however, cancels out the anomalies \cite{Garcia-Etxebarria:2018ajm}. 
The anomaly-free nature might be understood by embedding the $Z_4$ into the well-known $U(1)_{B-L}$ gauge group, but it is not necessarily. 
It might act on some other sectors like the inflation sector. 
It is very much welcome since we can control the inflaton potential to generate consistent inflation by the discrete symmetry. 
In fact, we constructed a topological inflation model based on the discrete $Z_4$ symmetry which is consistent with all present observations~\cite{Kawasaki:2023mjm}. 
However, this model predicts the Hubble parameter during inflation, $H_\mathrm{inf} \simeq 10^{12}$ GeV which causes too large isocurvature fluctuations of the axion dark matter.

In this paper, we extend the inflaton potential in a more general form allowed by the discrete $Z_4$ symmetry and show that the consistent hill-top inflation model is constructed. We find that the Hubble parameter $H_\mathrm{inf}$ can be smaller than $10^{9}$ GeV so that the axion isocurvature fluctuation is consistent with the present constraint. We also show that the reheating temperature can be as much as $T_R\simeq 10^{10}$ GeV and the thermal leptogenesis naturally works~\cite{Fukugita:1986hr}.

\section{Discrete \texorpdfstring{$Z_4$}{Z4} gauge symmetry in the standard model}
\label{sec:discrete_Z4}

We introduce a discrete $Z_4$ gauge symmetry in the standard model (SM). The charges are shown in Table~\ref{table:Z4_charge}. Here, we use the $SU(5)$ representations for the SM particles for simplicity of the notations, but we consider its subgroup $SU(3)\times SU(2)\times U(1)$ as a gauge group. It is stressed \cite{Garcia-Etxebarria:2018ajm} that the SM with the $Z_4$ is Dai-Freed anomaly free if we introduce one right-handed neutrino, $N_i$, for each generation $i=1,2,3$. Thus, we assume the SM with three right-handed neutrinos. 
It might be amusing that three right-handed neutrinos are required by the cancellation of the Dai-Freed anomalies in the SM with the gauged $Z_4$ symmetry.

We add a SM gauge singlet scalar $\phi$ to generate Majorana masses for the right-handed neutrinos, $N_i$. We consider a coupling of the $\phi$ to $N_i N_i$ as
\begin{equation}
    \label{eq:inflaton-Rneutrino}
    L=\frac{y_i}{2}\phi N_iN_i~ +~ h.c..,
\end{equation}
with $y_i$ coupling constants.
Here, the scalar $\phi$ should have the $Z_4$ charge ${2}$ mod 4, since $N_i$ have the  $Z_4$ charge 1 (see Table 1). This new scalar $\phi$ of the $Z_4$ charge 2 does not generate any Dai-Freed anomalies. 

\begin{table}[t]
    \centering
    \begin{tabular}{|c|ccccc|}
        \hline
               & $~5^*~$ & $~10~$   & $~N~$  & $~H~$  & $~\phi~$ \\
        \hline\hline
         $~Z_4~$ &  $1$  & $1$    & $1$  & $2$  & $2$   \\
        \hline 
    \end{tabular}
    \caption{$Z_4$ charges. $H$ is the Higgs doublet boson in the SM.}
    \label{table:Z4_charge}
\end{table}

\section{Hilltop inflation}
\label{sec:hilltop_inf}

In this paper, we consider the scalar $\phi$ as the inflaton. Notice that the inflaton sector has the discrete $Z_4$ symmetry common to the SM sector. This is very important, otherwise the inflaton decay to the SM particles is suppressed and the reheating temperature becomes too low to generate the baryon asymmetry.

Since the inflaton potential should be bounded from below and have the vanishing cosmological constant, we adopt the following $Z_4$ invariant form of the potential:
\begin{align}
    \label{eq: inf_pot_generic}
    V & =v^4 (1- \sum_{n=1} c_{2n}\phi^{2n} )^2.
\end{align}
Here and hereafter we use the Planck unit with the Planck mass $M_p = 2.4\times 10^{18}~\mathrm{GeV} = 1$.
In the previous paper~\cite{Kawasaki:2023mjm} we only considered the lowest-order term $c_2\phi^2$ and showed that topological inflation is realized.
In this paper, we consider a more generic potential,
\begin{align}
    \label{eq:inf_pot_c}
    V & =v^4 (1- c_2 \phi^2 - c_4 \phi^4 -c_6 \phi^6 )^2,
\end{align}
where we neglect higher order terms ($\phi^8, \ldots$) in Eq.~\eqref{eq: inf_pot_generic}.
For the inflaton not to be trapped at $\phi=0$, $c_2$ should be positive.
We further assume that $c_6> 0$ and $|c_4|/c_6^{2/3} \ll 1$.
The vacuum expectation value of the inflation is then given by $\langle \phi\rangle\equiv M = c_6^{-1/6}$.
Using $M$, we rewrite the potential~\eqref{eq: inf_pot_generic} as
\begin{align}
    \label{eq:inf_pot}
    V & =v^4 \left(1- \alpha \frac{\phi^2}{M^2} 
    - \beta \frac{\phi^4}{M^4} -\frac{\phi^6}{M^6} \right)^2 
    \nonumber\\
    & =  v^4 (1-\alpha \varphi^2 
    - \beta \varphi^4 -\varphi^6)^2
    \nonumber\\
    & \equiv v^4 U(\varphi),
\end{align}
where $\alpha = c_2 M^2$, $\beta = c_4 M^4$ and $\varphi = \phi/M$.

Let us study the dynamics of the inflaton.
The slow-roll parameters are written as 
\begin{align}
    \epsilon & = \frac{1}{2M^2}\left(\frac{U'(\varphi)}{U(\varphi)}\right)^2 ,\\
    \eta & = \frac{1}{M^2}\frac{U''(\varphi)}{U(\varphi)} ,\\
    \label{eq:xi}
    \xi & = \frac{1}{2M^4}
        \frac{U'''(\varphi)U'(\varphi)}{U(\varphi)^2},
\end{align}
where $' \equiv d/d\varphi$.
Sufficient inflation takes place when the slow-roll parameters are much smaller than 1, and inflation ends at $t\simeq t_\mathrm{f}$ when $|\eta | $ or $\epsilon$ becomes $\simeq 1$.
Since $\epsilon \ll \eta$ for $M  \ll 1$, the field value $\phi_\mathrm{f} = M\varphi_\mathrm{f}$ at the end of inflation is determined by $|\eta | \simeq 1$.
The e-fold number $N (\,=\ln [ a(t_\mathrm{f})/a(t_N)]\, )$ is given by
\begin{align}
    \label{eq:n-efold}
    N & = \int_{\phi_\mathrm{f}}^{\phi_N}d\phi\,
    \frac{1}{\sqrt{2\epsilon}}
    = M^2 \int_{\varphi_\mathrm{f}}^{\varphi_N}d\varphi\,
    \frac{U(\varphi)}{U'(\varphi)},
\end{align}
where $\phi_N =M \varphi_N$ is the inflaton field value at $t=t_N$.

\begin{figure}[t]
    \centering
    \includegraphics[width=0.52\textwidth]{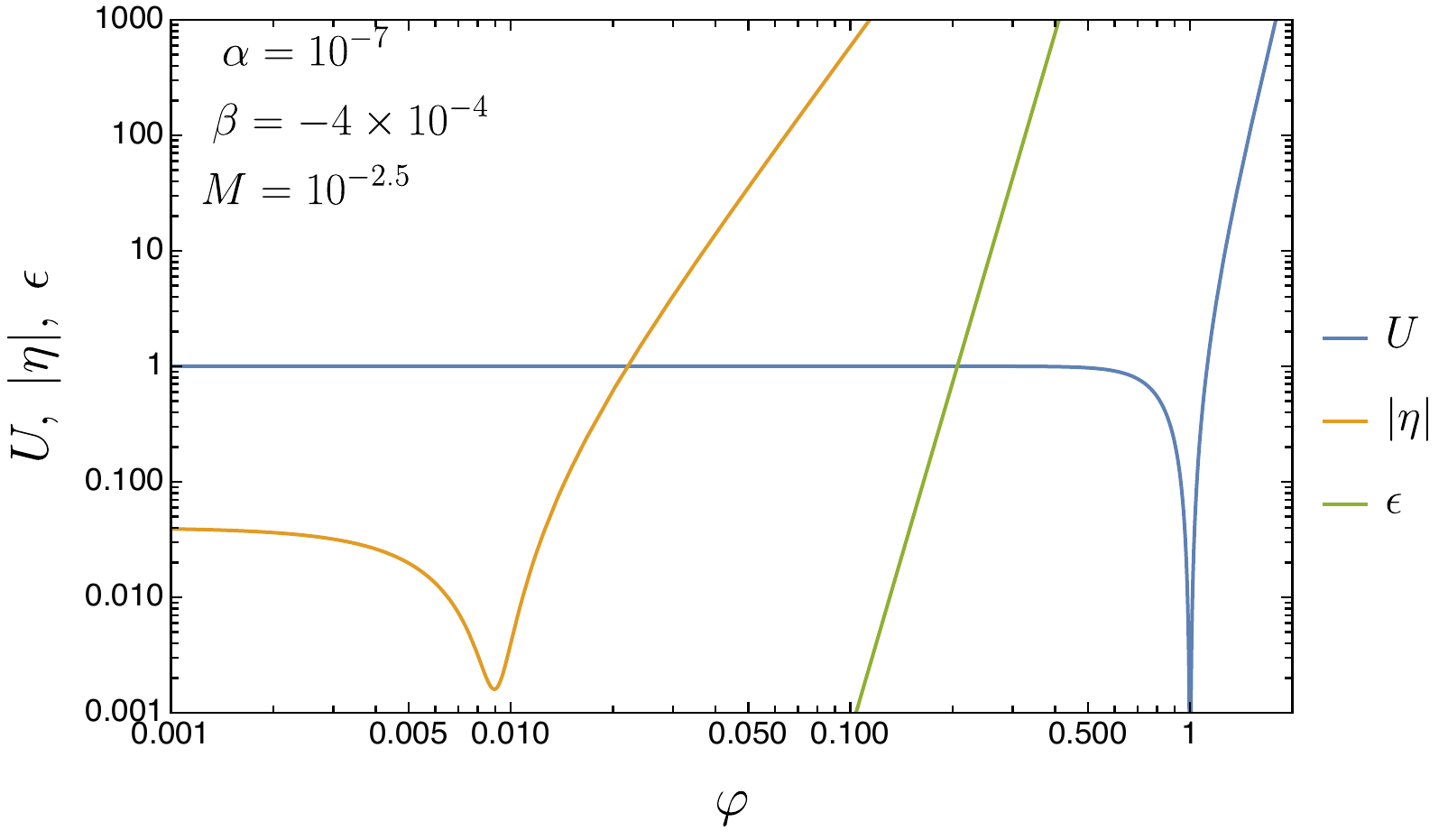}
    \includegraphics[width=0.47\textwidth]{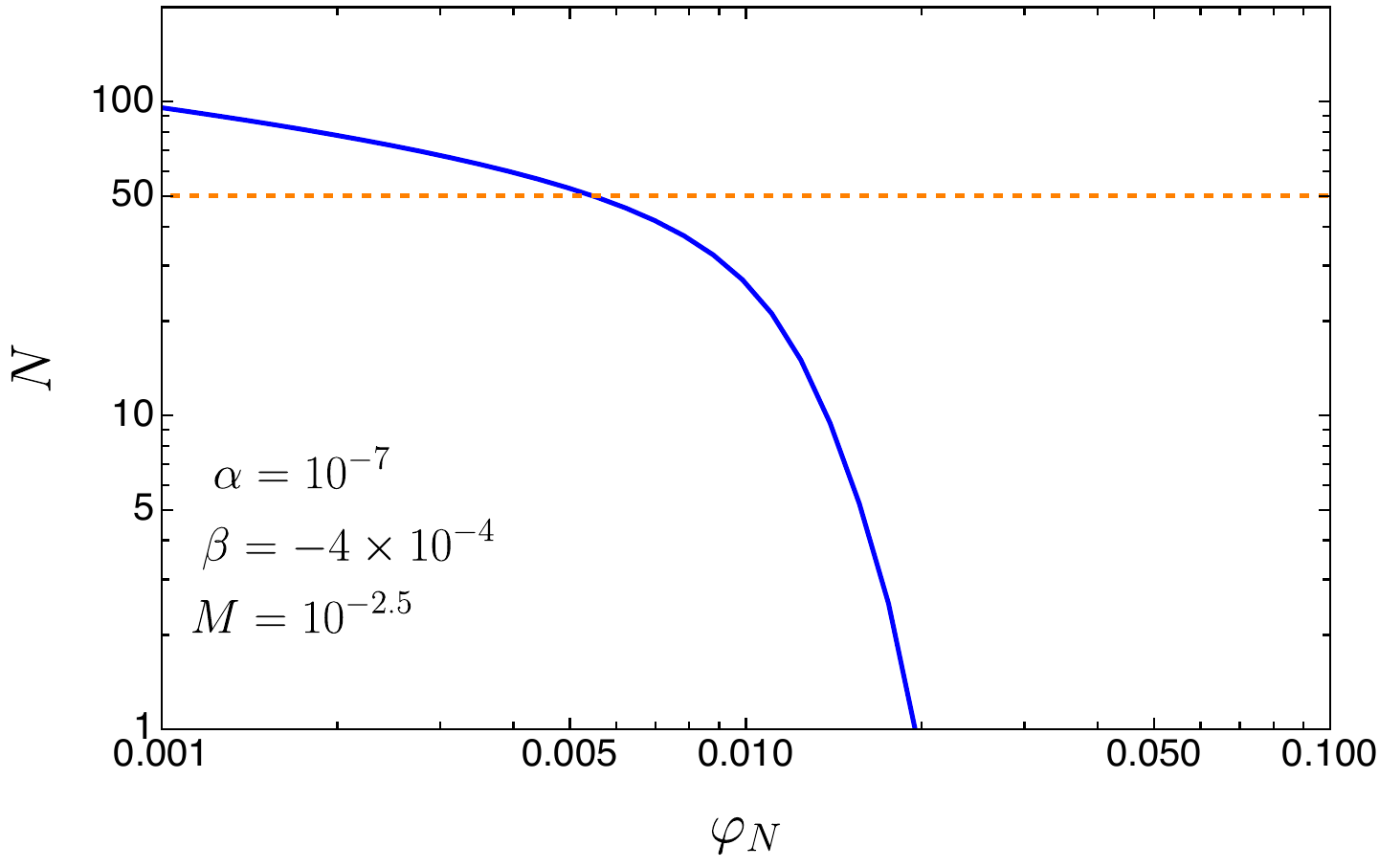}
    \caption{%
        Potential $U(\varphi)$ and slow-roll parameters ($\epsilon$ and $\eta$) are shown as a function of $\varphi=\phi/M$ for $\alpha=10^{-7}$, $\beta=-4\times 10^{-4}$ and $M=10^{-2.5}$ in the left panel.
        In the right panel, we show the relation between the e-fold number $N$ and $\varphi_N$ by the blue line.
        The dashed orange line represents $N=50$.
        }
    \label{fig:potential-efold}
\end{figure}

Using the slow-roll parameters, the spectral index $n_s$ of the curvature perturbation, the tensor-to-scalar ratio $r$ and the running of the spectral index $dn_s/d\ln k$ are written as
\begin{align}
    n_s & = 1-6\epsilon + 2\eta, 
    \label{eq:spectral_index}\\
    r & = 16\epsilon,\\
    \frac{dn_s}{d\ln k} & = 16\epsilon\eta -24\epsilon^2 -2\xi,
\end{align}
where $k$ is the wavenumber of the curvature perturbation.
We obtain $\varphi_\mathrm{f}$ by numerically solving the equation $|\eta|=1$.
The inflaton field value when the CMB scale ($k_*=0.05\,\mathrm{Mpc}^{-1}$) exits the horizon during inflation is then calculated by solving Eq.~\eqref{eq:n-efold} with $N=N_* \in [50,60]$.
We show the potential $U(\varphi)$ and slow-roll parameters ($\epsilon$ and $\eta$ ) in Fig.~\ref{fig:potential-efold} for $\alpha=10^{-7}$, $\beta=-4\times 10^{-4}$ and $M=10^{-2.5}(\,\simeq 7.6\times 10^{15}\,\text{GeV})$. 
It is seen that $\epsilon$ is much smaller than $|\eta|$ and $|\eta|$ has a minimum at $\varphi \sim 0.1$ due to negative $\beta$ (see Sec.~\ref{subsec:NIP} for more details).
In Fig.~\ref{fig:potential-efold} we also show the relation between $\varphi_N$ and $N$ obtained from Eq.~\eqref{eq:n-efold}.

The amplitude of the power spectrum of the curvature perturbations at the CMB scale is written as
\begin{align}
    \label{eq:inf_scale}
    \mathcal{P}_\zeta (k_*) & = \frac{1}{24\pi^2}\frac{V}{\epsilon} 
     = \frac{v^4 M^2}{12\pi^2}\frac{(U(\varphi_{N_*}))^3}{(U'(\varphi_{N_*}))^2}.  
\end{align}
Using the observed amplitude $\mathcal{P}_\zeta(k_*) = 2.1\times 10^{9}$~\cite{Planck:2018vyg}, we evaluate the inflation scale $v$.
The Hubble parameter during inflation $H_\mathrm{inf}$ is then obtained as
\begin{equation}
    \label{eq:hubble}
    H_\mathrm{inf} \simeq \frac{v^2}{\sqrt{3}}.
\end{equation}
The inflaton mass $m_\phi$ is also written as
\begin{equation}
    \label{eq:inflaton_mass}
    m_\phi = 6\sqrt{2}\,\frac{v^2}{M} 
    = 6\sqrt{6}\,\frac{H_\mathrm{inf}}{M} .
\end{equation}

\begin{figure}[t]
    \centering
    \includegraphics[width=0.7\textwidth]{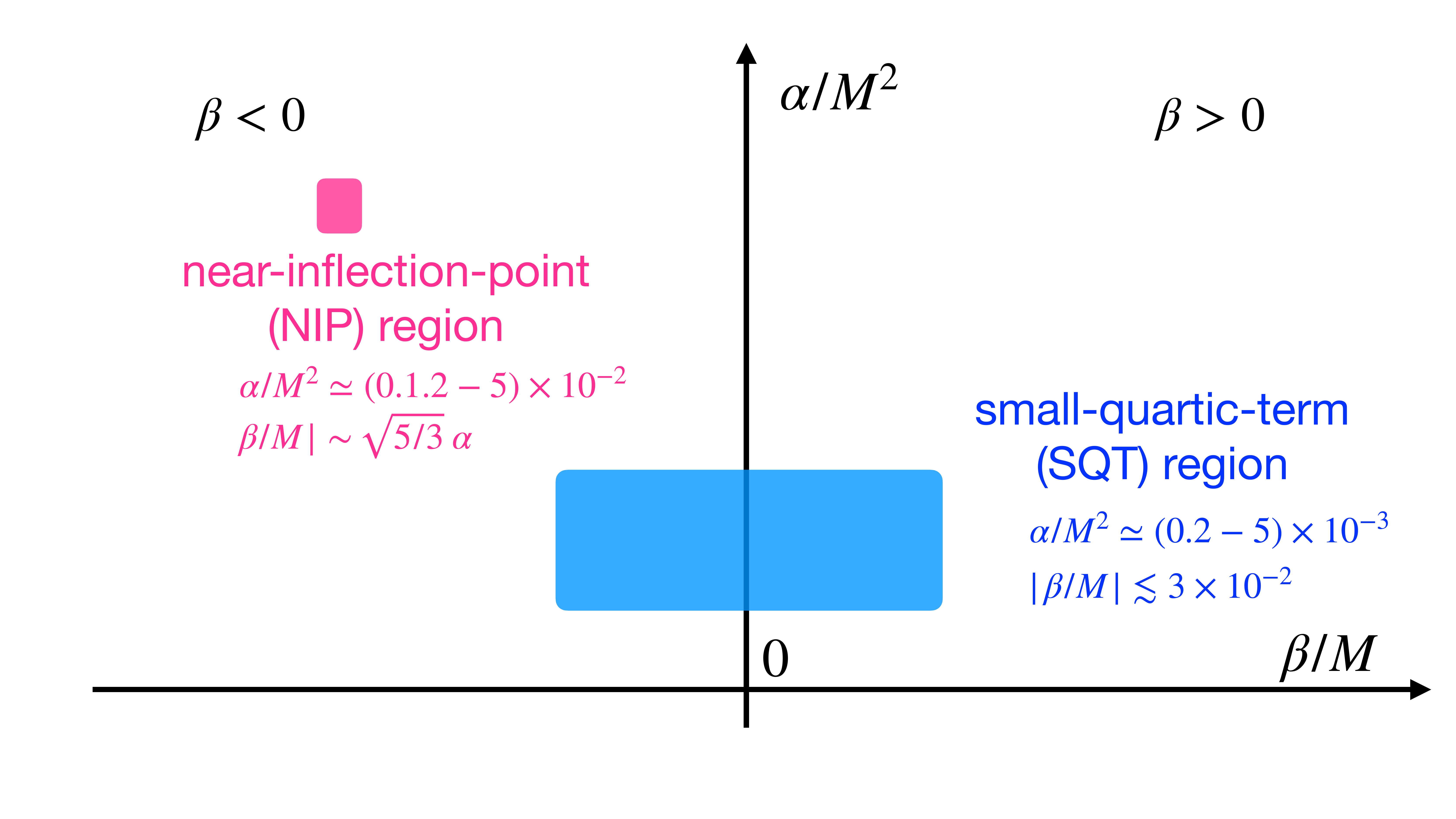}
    \caption{%
        Schematic picture showing two model parameter regions.}
    \label{fig:successful_region}
\end{figure}

Successful hilltop inflation takes place for two model parameter regions.
One is the small-quartic-term region (SQT region) and the other is the near-inflection-point region (NIP region).
Fig.~\ref{fig:successful_region} illustrates schematically these regions whose details are described below. 

\subsection{Small-quartic-term region region}

\begin{figure}[t]
    \centering
    \includegraphics[width=0.49\textwidth]{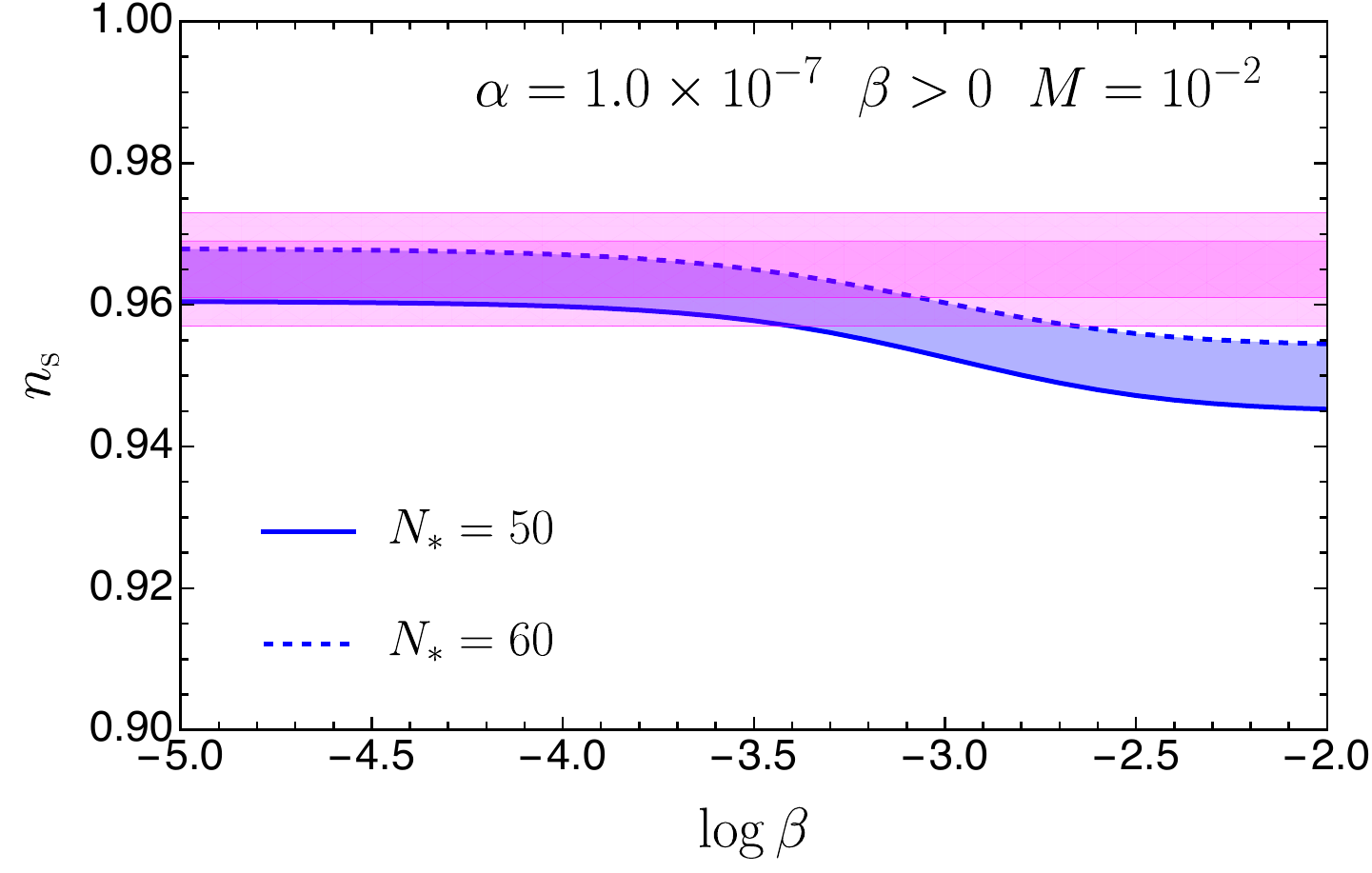}
    \includegraphics[width=0.49\textwidth]{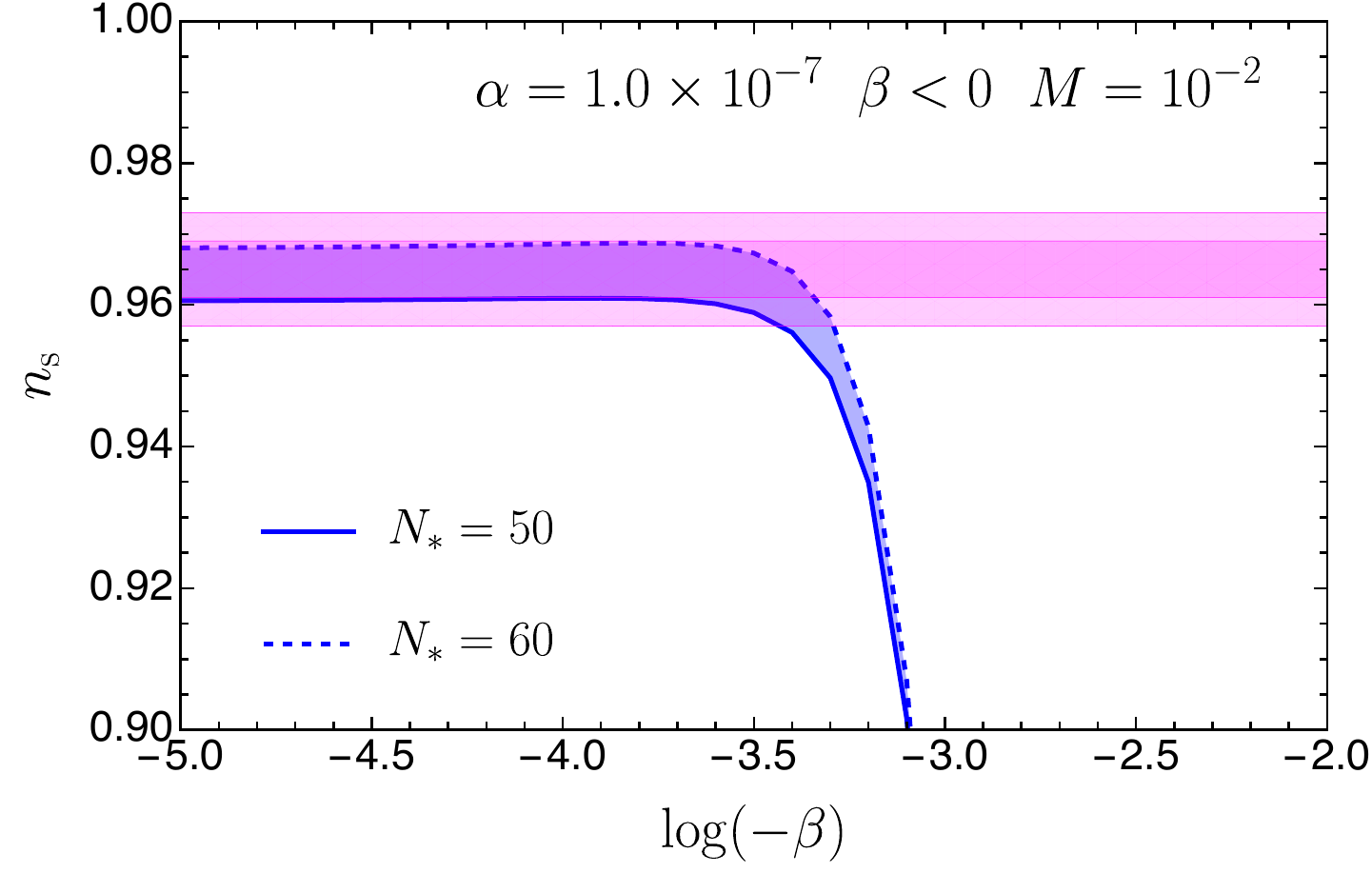}
    \caption{%
        Dependence of the spectral index on  $\beta$ for  $\alpha=10^{-7}$ and $M=10^{-2}$.
        The solid (dashed) line represents the spectral index for $N_*=50 \,(60)$.
        The dark (light) pink-shaded region shows the allowed $1 \sigma$ ($2 \sigma$) range of the spectral index from Planck~2018~\cite{Planck:2018vyg}.
        The left (right) panel show the case of $\beta >0$ ($\beta <0$).}
    \label{fig:specindex_SQT}
\end{figure}

\begin{figure}[t]
    \centering
    \includegraphics[width=0.75\textwidth]{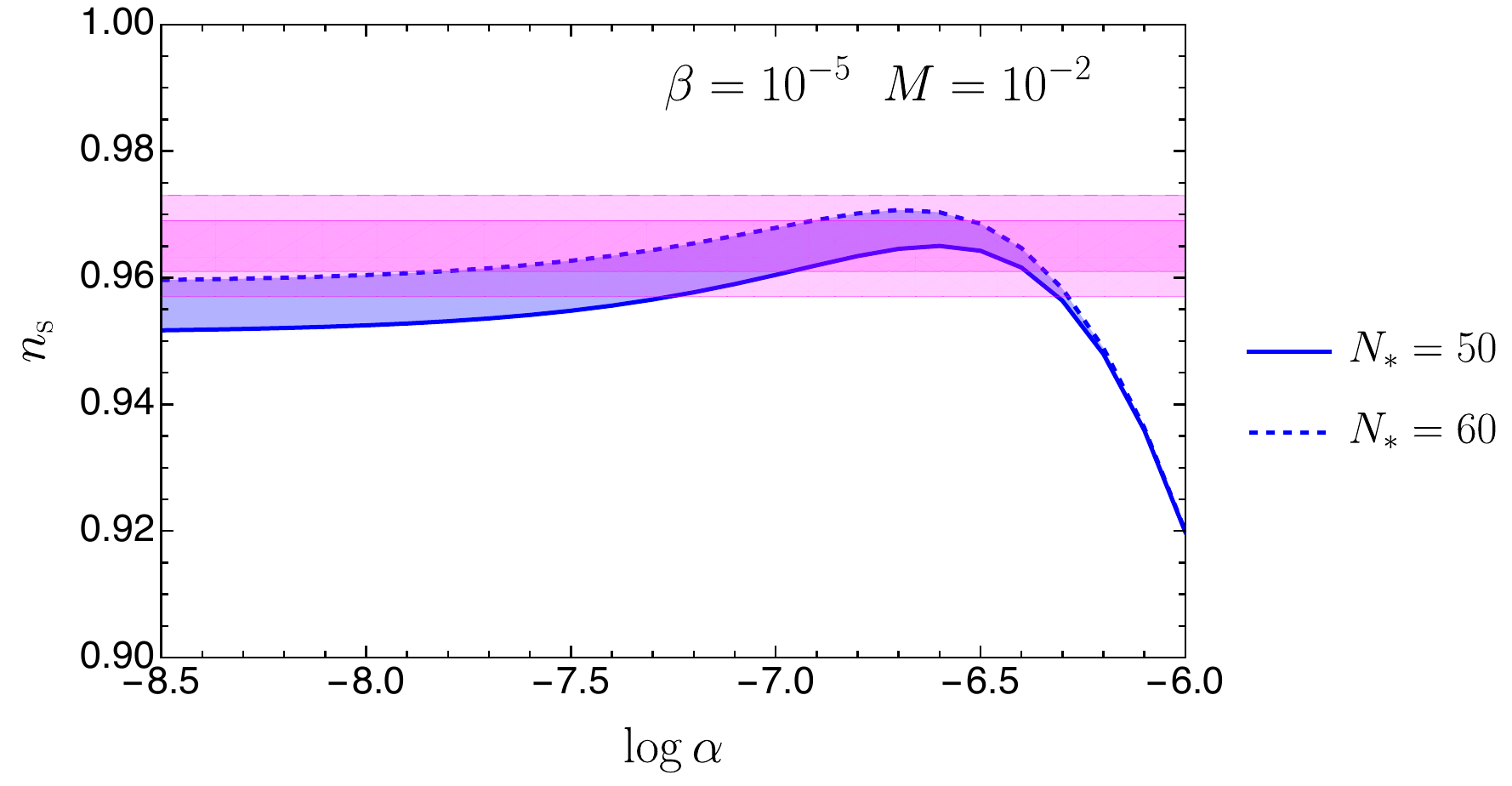}
    \caption{%
        Dependence of the spectral index on $\alpha$ for  $\beta=10^{-5}$ and $M=10^{-2}$.}
    \label{fig:specindex_alpha}
\end{figure}

In SQT region, the inflaton dynamics is governed by the $\phi^6$ and $\varphi^2$ terms, and the potential is approximated as $U \simeq (1 -\alpha\varphi^2-\varphi^6)^2\simeq 1-2\alpha \varphi^2-2\varphi^6$.
We show the spectral index~Eq.~\eqref{eq:spectral_index} for $\alpha=10^{-7}$ and $M=10^{-2}$ in Fig.~\ref{fig:specindex_SQT} together with the observational constraint $n_s = (0.965\pm 0.004\,(1\sigma))$ given by Planck~2018~\cite{Planck:2018vyg}.
One can see that the predicted spectral index is consistent with the observation for sufficiently small $\beta$.
We investigate the constraint on $\beta$ and its dependence on $M$ from the observation, and find $|\beta | \lesssim 3\times 10^{-2}M$.
The spectral index also depends on $\alpha$ as shown in Fig.~\ref{fig:specindex_alpha} and it is found that the observationally consistent spectral index is obtained for $\alpha  \simeq (0.2-5)\times 10^{-3}M^2$.
Thus, the SQT region is given by
\begin{align}
    \alpha & \simeq (0.2-5)\times 10^{-3}M^2, \\[0.5em]
    |\beta | & \lesssim 3\times 10^{-2}M.
\end{align}


The result for the SQT region agrees with the previous work on SUSY hilltop inflation~\cite{Harigaya:2013pla}.
In Ref.~\cite{Harigaya:2013pla} a SUSY model gives a potential $V\simeq v^4 -\kappa v^4\varphi^2 - gv^2 \varphi^n$ ($n=4,5,6$) with $\kappa$ and $g$ the coupling constants, and inflation consistent with the observation occurs for $n=6$.

The Hubble parameter during inflation and inflation mass in the SQT region are calculated from Eqs.~\eqref{eq:inf_scale}, \eqref{eq:hubble} and \eqref{eq:inflaton_mass} and we find that they are given by 
\begin{align}
    \label{eq:hubble_SQT}
    H_\mathrm{inf} & \simeq (4-9)\times 10^{8}\,\mathrm{GeV} \left(\frac{M}{0.01}\right)^{3/2}, \\
    \label{eq:inflaton_mass_SQT}
    m_\phi & \simeq (0.6-1)\times 10^{12}\, \mathrm{GeV}\,
    \left(\frac{M}{0.01}\right)^{1/2}.    
\end{align}
Thus, the model predicts a small Hubble parameter during inflation, which is favored by axion dark matter.
One serious problem of the axion dark matter is that it has too large isocurvature perturbations unless $H_\mathrm{inf}$ is sufficiently small~(e.g. \cite{Kawasaki:2013ae}).
In particular, for string-theory inspired axion with axion decay constant $F_a \simeq 10^{16}\,\mathrm{GeV}$, the isocurvature constraint is $H_\mathrm{inf} \lesssim 10^9$~GeV~\cite{Kawasaki:2015pva}.
From Eq.~\eqref{eq:hubble_SQT} the hilltop inflation model avoids the isocurvature-perturbation problem if $M \lesssim  10^{-2} \simeq 2.4 \times 10^{16}$~GeV.

Since the Hubble parameter is low, the tensor mode is hardly generated.
The tensor-to-scalar ratio $r$ is $r \simeq (0.6-1)\times 10^{-11}(M/0.01)^3$ in the SQT region. 
Thus, even if we take $M=1$, $r \lesssim 10^{-5}$ which is much smaller than the present observational constraint $r < 0.036\,(2\sigma)$~\cite{BICEP:2021xfz}.

\subsection{Near-inflection-point region}
\label{subsec:NIP}

Successful inflation also takes place around the NIP region which is characterized by  a sharp decrease of $\eta $ around the CMB scale. 
This occurs when $\beta < 0$ and $U''$ is nearly zero at some $\varphi$.
In fact, an example of this situation is seen in Fig.~\ref{fig:potential-efold} where one can see that $|\eta|$ drops sharply at $\varphi \sim \varphi_{N_*}\sim \varphi_{N_{50}} \sim 0.01$.
In the inflationary region ($\varphi \ll 1$) the potential $U$ and $\eta$ are approximately given by 
\begin{align}
    U & \simeq 1-2\alpha \varphi^2 -\beta\varphi^4 -\varphi^6 ,\\
    \eta & \simeq \frac{U''}{M^2}\simeq \frac{1}{M^2}
         \left(-4\alpha -24\beta \varphi^2 -60\varphi^4\right),
\end{align}
from which the $\eta$ has the minimum  $-4(\alpha - 3\beta^2/5)$ at $\varphi_\mathrm{min} \simeq \sqrt{-\beta/5}$.
Thus, $\eta(\varphi_\mathrm{min}) \simeq U''(\varphi_\mathrm{min})/M^2\sim  0$ if $\alpha \sim 3\beta^2/5$, and $\eta$ drops sharply there. 

\begin{figure}[t]
    \centering
    \includegraphics[width=0.75\textwidth]{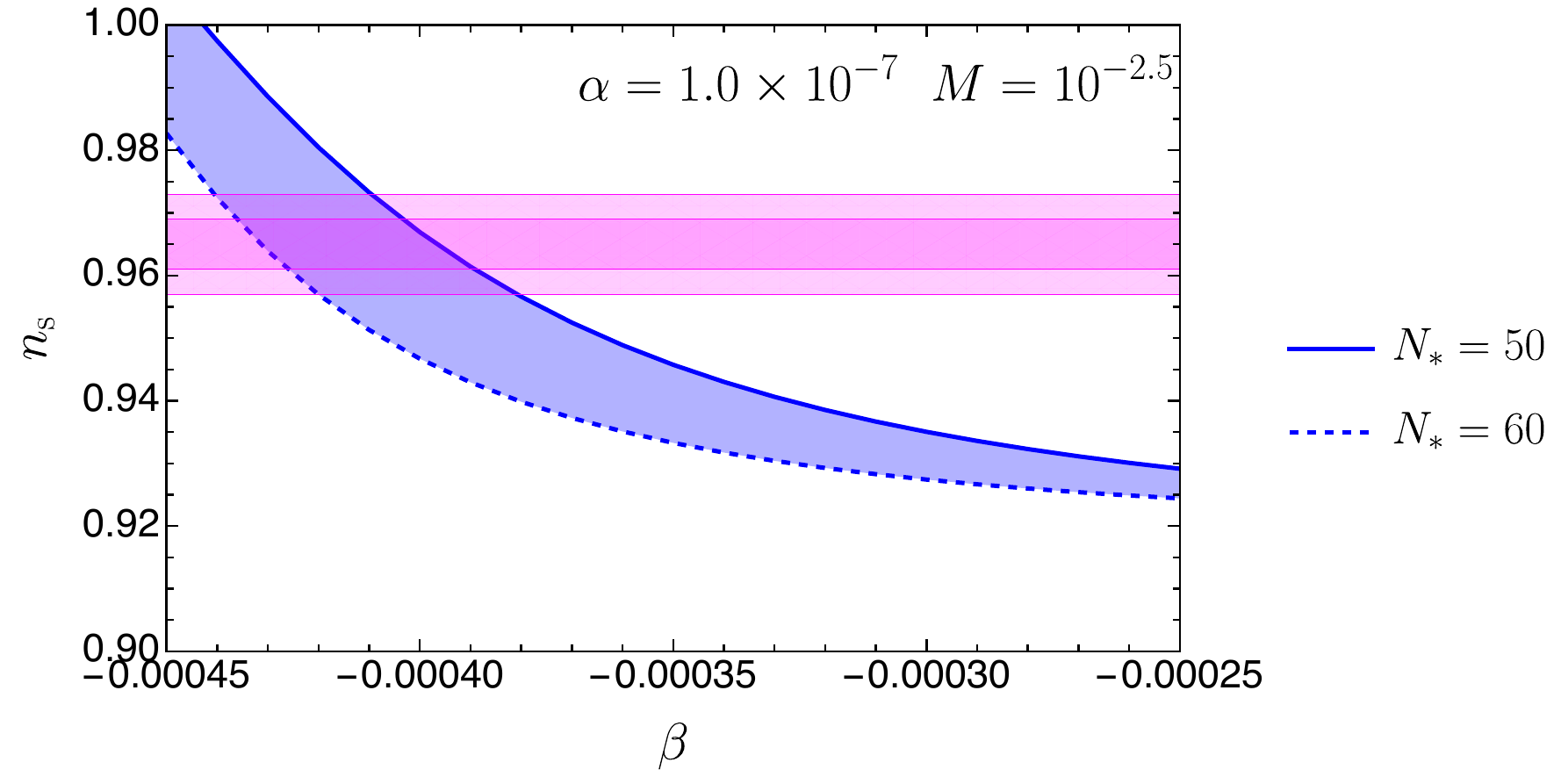}
    \caption{%
        Spectral index in the near-inflection-point region.
        We take $\alpha=10^{-7}$ and $M=10^{-2.5}$.
        }
    \label{fig:specindex_NIP}
\end{figure}

In Fig.~\ref{fig:specindex_NIP} we show the spectral index around the NIP region for $\alpha = 10^{-7}$ and $M=10^{-2.5}$, for which the inflection point $U''\simeq 0$ is realized for $\beta_\mathrm{IP} \simeq 4.2\times 10^{-4}$.
In this case, from Eq.~\eqref{eq:spectral_index}, the spectral index is given by $n_s \simeq 1+2\eta \simeq 1-8\alpha M^{-2} = 0.92$ for $\beta$ large than $\beta_\mathrm{IP}$.
However, as $\beta$  approaches $\beta_\mathrm{IP}$ the spectral index increases because $\eta \simeq M^{-2}U''$ increases from negative to positive values.
The specral index consistent with the CMB observation is obtained near $\beta_\mathrm{IP}$, i.e. $\beta\simeq -(4.1\pm 0.2)\times 10^{-4}$.
From the above argument and numerical calculations, the NIP region for successful hilltop inflation is characterized by
\begin{align}
    \alpha & \simeq (0.8-1.2)\times 10^{-2}M^2 ,\\
    \beta & \sim -\sqrt{\frac{5}{3}\alpha} .
\end{align}

It is also noticed that the spectral index is sensitive to $N_*$ in comparison to the SQT case.
This implies that the running of the spectral index is large in the NIP region.
For example, for $\alpha=1.2\times 10^{-7}$, $\beta=4.8\times 10^{-4}$, $M=10^{-2.5}$ and $N_*=50$, we obtain the spectral index and its running as
\begin{align}
    n_s & = 0.964 , \\
    \frac{dn_s}{d\ln k} & = 0.0018 .
\end{align}
The large spectral index is explained by the sharp change of $\eta =U''/M^2$, which leads to large $\xi \propto U'''$ and hence large $dn_s/d\ln k$ [see, Eq.~\eqref{eq:xi}].
The present observational constraint on the spectral index running is $dn_s/d\ln k = -0.0045\pm 0.0067 (1\sigma)$~\cite{Planck:2018jri}.
Thus, the predicted spectral index running is consistent with the observation and will be tested in the future. 

The Hubble parameter and inflaton mass in the NIP region are given by
\begin{align}
    H_\mathrm{inf} 
    & \simeq 2\times 10^9\, \mathrm{GeV}
    \left( \frac{M}{0.01} \right)^{3/2}, \\
    \label{eq:inflaton_mass_NIP}
    m_\phi 
    & \simeq 3\times 10^{12}\, \mathrm{GeV}\,
    \left(\frac{M}{0.01}\right)^{1/2}.
\end{align}
Thus, the predicted Hubble parameter and inflaton mass are larger than those in the SQT region.
In order to avoid the isocurvature perturbation problem in the string axion the inflaton vacuum expectation value $M$ should be smaller than about $1.5\times 10^{16}$~GeV.
On the other hand, the tensor-to-scalar ratio is $r\simeq 7\times 10^{-11}(M/0.01)^3$ which is below  experimental reach for $M<1$. It is remarkable that if the tensor-to-scalar ratio $r$ is measured  in future CMB experiments we can determine the effective cut-off scale $M$.

\section{Reheating temperature and leptogenesis}
\label{sec:leptogenesis}

The inflaton $\phi$ decays into the right-handed neutrinos through the interaction~\eqref{eq:inflaton-Rneutrino}.
The decay occurs when the right-handed neutrino mass $m_{N_i}\, (= y_i M)$ is less than $m_\phi/2$.
Here, for simplicity, we assume $m_{N_3}  > m_\phi/2 > m_{N_2}\gg m_{N_1}$.
The inflaton then decays mainly into $N_{2}$'s and the decay rate of the inflaton is written as
\begin{align}
    \Gamma_\phi & = \frac{1}{8\pi} y_2^2  m_\phi 
    = \frac{1}{8\pi}\frac{m_{N_2}^2}{M^2}m_\phi,
\end{align}
from which the reheating temperature is estimated as 
\begin{align}
    T_R & \simeq \left(\frac{\pi^2 g_*}{90}\right)^{1/4}
    \Gamma_\phi^{1/2}
    = \frac{1}{4}\left(\frac{2 g_*}{45}\right)^{1/4}
    \frac{m_{N_2}}{M}m_\phi^{1/2} ,
\end{align}
where $g_*$ is the relativistic degree of freedom.
Using the relation between $m_\phi$ and $M$ [Eqs.~\eqref{eq:inflaton_mass_SQT} and \eqref{eq:inflaton_mass_NIP}] we obtain
\begin{equation}
    \label{eq:reheating_temp}
    T_R \simeq 
    \begin{cases}
    \begin{aligned}
        2\times 10^{9}~\mathrm{GeV}&
        \left(\frac{m_{N_2}}{3 \times 10^{11}\mathrm{GeV}}\right)
        \left(\frac{M}{0.01}\right)^{-3/4}
        \left(\frac{g_*}{100}\right)^{-1/4}
        & ~~~~(\mathrm{SQT}) ,\\[0.5em]
        1.2\times 10^{10}~\mathrm{GeV}&
        \left(\frac{m_{N_2}}{10^{12}\mathrm{GeV}}\right)
        \left(\frac{M}{0.01}\right)^{-3/4}
        \left(\frac{g_*}{100}\right)^{-1/4}
        & ~~~~(\mathrm{NIP}).
    \end{aligned}    
    \end{cases}
\end{equation}

Since we have three species of right-handed neutrinos in this model, it is natural to consider that the baryon number in the universe is generated through leptogenesis due to the right-handed neutrino decay.
For successful thermal leptogenesis, the reheating temperature should be $T_R \gtrsim 1.9\times 10^9$~GeV for the lightest right-handed neutrino mass $m_{N_1} \sim 10^{9}$~GeV~\cite{Antusch:2006gy}.
Eq.~\eqref{eq:reheating_temp} implies that the reheating temperature in both successful parameter regions with $M \lesssim 10^{-2}$ is high enough for thermal leptogenesis. 

The maximum reheating temperature is obtained when the right-handed neutrino mass $m_{N_2}$ is near to $m_\phi/2$.
Since the inflaton mass depends on $M^{1/2}$, the maximum reheating temperature scales as $M^{-1/4}$.
Thus, the reheating temperature is lower than $10^9~\mathrm{GeV}$ for $M\gtrsim 0.1$ in the SQT region.
On the other hand, in the NIP region, the high reheating temperature $10^9$~GeV is achieved even for $M=1$.

\section{Conclusions}
\label{sec:Conclusions}

Motivated by the Dai--Freed anomaly cancellation in the standard model with three right-handed neutrinos we consider the discrete $Z_4$ gauge symmetry. 
A scalar field $\phi$ is required to generate the Majorana masses for the right-handed neutrinos and hence this scalar field carries the $Z_4$ charge 2. 
We identify this scalar $\phi$ with the inflaton, and we can control its potential by the discrete symmetry $Z_4$. 
The identification of the discrete symmetries acting on the inflaton and the SM sector has a remarkable merit that the inflaton can couple directly to the right-handed neutrino generating fast inflaton decays. 
As a consequence we have a relatively high reheating temperature $T_R \simeq 10^{9-10}$ GeV and the thermal leptogenesis naturally works.

In this paper we have constructed a hilltop inflation model consistent with all present observations. 
In particular, we have shown that the Hubble parameter during inflation can be made as $H_\mathrm{inf} \lesssim 10^{9}$ GeV to suppress the axion isocurvature fluctuations below the observational constraints for the string-theory inspired axion model with the decay constant $f_a \simeq 10^{16}$ GeV. 
It is very difficult to discover the tensor modes in future CMB experiments since the inflation scale is too low as $H_\mathrm{inf} \lesssim 10^{9}$ GeV.
We have, however, found that the running of the spectral index can be significantly large, $dn_s/\ln k  \simeq  0.0018$ which will be tested in the future CMB observations.

 \section*{Acknowledgements}

T. T. Y. thanks Kazuya Yonekura for the discussion on the Dai-Freed anomaly. T. T. Y. is supported by the China Grant for
Talent Scientific Start-Up Project and by Natural Science
Foundation of China (NSFC) under grant No.\,12175134,
JSPS Grant-in-Aid for Scientific Research
Grants No.\,19H05810, 
and World Premier International Research Center
Initiative (WPI Initiative), MEXT, Japan.
M. K. is supported by JSPS KAKENHI Grant Nos. 20H05851(M.K.) and 21K03567(M.K.). 

\bibliographystyle{JHEP}   
\bibliography{Ref}

\end{document}